\def\agt{\,\raise.3ex\hbox{$>$\kern-.75em\lower1ex\hbox{$\sim$}}\,}
\def\alt{\,\raise.3ex\hbox{$<$\kern-.75em\lower1ex\hbox{$\sim$}}\,}

\documentstyle[epsfig]{mnkop}

\title{Microlensing, structure of the galactic halo and
 determination of dark objects' mass function} 
\author[D. Markovi\'{c} and J. Sommer-Larsen]{D. Markovi\'{c} and 
    J. Sommer-Larsen\\
    Theoretical Astrophysics Center, Juliane Maries Vej 30, DK-2100
    Copenhagen \O, Denmark}
\begin{document}
\input{epsf}

\maketitle

\begin{abstract}
We study the accuracy and systematic error of inference
of massive halo objects' (MHO or `Macho') mass function from microlensing
events observed in the direction of Large Magellanic Cloud.  Assuming the spatial 
distribution and kinematics of the objects are  known, the slope and the  range
of the MHO mass function (modeled here by a simple power law)
will be possible to determine from 100-1000 detected events {\it if}
the slope is in the range $-2.5\alt \alpha\alt -0.5$, with the statistical
errors reaching their minima at $\alpha =-1.5$.
Outside this range the  errors grow rapidly making the inference
difficult even at very large numbers of events ($N\approx 10000$).
On the other hand, the {\it average} mass of the MHO's will 
be determined to better than about 30\% accuracy from $N\approx 100$ events 
for {\it any} slope.
Overall, we find that the accuracy of inference at fixed $N$ will not be 
strongly affected by the presently
available event duration-dependent detection efficiencies if the typical
MHO masses are in the range (order of magnitude $ 0.1 M_{\odot}$) indicated 
by the events detected so far. 

We also estimate the effects of the uncertainty of the massive objects'
spatial distribution and kinematics on the determination of their mass function.
The massive objects' halo models considered are all spherical   but we allow for
various density profiles and a radius-dependent, anisotropic velocity dispersion. 
We find that while the mass function slope and range (i.e.,
the `shape')  are weakly affected for $-2 \alt\alpha\alt 0$, the error in
the average mass due to the halo structure uncertainties could be reduced to less than 
about 50\% only through the detection of about 1000 or more events.
Reliable estimates of the
halo structure itself [density profile and (anisotropic) velocity dispersion
profile] can start only at very large numbers of detections ($N \agt 10000$).

\end{abstract}


\section{Introduction and overview}
A decade after Paczynski's ground-breaking proposal (1986) several teams and
considerable resources around the world are now devoted to searches 
for microlensing events [see recent overviews by Paczynski (1996) and
Roulet \& Mollerah (1996)].  These searches have been rewarded by 
by a large number ($\sim 100$) of
detections in the direction of the galactic bulge and already
a significant set of events observed [1 or 2 by the EROS team \cite{aubourg.93}
and 6-8 by the MACHO team \cite{alcock.96}] along the line of sight to the
Large Magellanic Cloud.   

In the latter case microlensing is expected to provide
us with direct information on the composition of the galactic halo. A statistical
analysis \cite{alcock.96} of the presently available
data suggests that dark massive objects --- the potential microlenses ---
could account for between 30\% and 100\% of the total
mass in the halo. In addition, their most likely masses should be
in the range 0.1-0.6$M_{\odot}$.  
The admittedly large uncertainty stems in part from the still
small number of events and thus a poor statistics, but also from
our lack of knowledge [as made clear in \cite{alcock.96}] regarding
the spatial distribution and kinematics of the massive halo objects (MHO). 
The import of the halo structure is due to the simple fact that
the (inferred) mass of a lens scales as $m \sim (T v_{\rm n})^2 /D_{\rm L}
(D_{\rm S} - D_{\rm L})$, where $T$ is the {\it observed}
duration of the microlensing event, $v_{\rm n}$ is the MHO's velocity
orthogonal to the line of sight, $D_{\rm L}$ is the earth-lens distance
and $D_{\rm S}$ is the earth-source distance.  At the same time, the integral 
event rate which is roughly proportional to 
$\int \rho \sqrt{D_{\rm L}(D_{\rm S} - D_{\rm L})}
v_{\rm n} d D_{\rm L}$ obviously incorporates details about halo  beyond
the mere total mass $M_{\rm tot} \sim \int\rho d^3 D_{\rm L}$.
Therefore, the conversion from the observed quantities, the event durations and
the rate, to information on MHOs' masses and their halo's total mass inevitably involves 
assumptions about the structure of the halo.

A question arises at this point: should we expect --- as might seem
only natural --- that, given a larger number of detections in a 
foreseeable future, the determination of the MHO
mass function and the MHOs'  fraction of the total halo mass could  reach
a significantly higher accuracy thus providing us with better
clues regarding the origin and evolution of the halo. 
 The present paper will address this question.

Recently Mao and Paczynski (1996) studied the issue
of MHOs' mass  determination.  By considering simplified
`toy' models (e.g., uniform spatial density of MHOs) they were
able to estimate
that a reliable determination of the mass function could be achieved
if we had 100 or more events.  They modeled the mass function
by a simple power law $dn/dm \propto m^{\alpha}$ (as we shall do in this 
paper) and concluded that for $\alpha \ll -1.5$ ($\alpha\gg -1.5$) the
high (low) mass end of the mass function will be difficult to probe.
As they pointed out, their results depended on the assumption that the MHOs'
spatial distribution and kinematics were known.

In this paper we also at first assume that the halo model is known.
The specific model used in Section 4  for the purpose of mass function 
inference is the isothermal sphere with a core (CIS). Although the underlying
halo model is considerably more realistic in our case, the basic conclusion of
Mao and Paczynski remains valid: in the vicinity of
$\alpha =-1.5$ the slope and range $\beta = \log_{10}(m_{\rm max}/m_{\rm min})$
can be determined with $N\agt 100$ events.  However, as we shift away from 
$\alpha =-1.5$,
the error of determination grows very rapidly.  In particular,
for a positive slope $\alpha$ one needs $N > 1000$ events for a reliable 
inference.  [It is to be hoped --- perhaps not unrealistically --- that
the actual mass function of the MHOs will indeed correspond to $\alpha$
sufficiently close to -1.5.]  A quantity that can be accurately ($\alt
30\% $ error)  inferred with $N\agt 100$ at any slope is the
average mass of the MHOs, i.e., the first moment of the mass function.
We find that our results do not depend dramatically (see, e.g., Fig. 6) on whether the
detection efficiency is flat (i.e. independent of event duration) or 
of the type presently available for the MACHO project's microlensing
searches \cite{alcock.96}.

The effects of halo structure uncertainty will receive our attention in the later
part (section 5) of the paper.  At the outset of section 5
 we will perform the following 
`experiment:' we will assume that a specific
spherical halo model describes the actual halo accurately enough (this
model will be called the `real' one).
 We will then ask what the effect on mass function and 
 MHOs' halo fraction determination is if, instead of the `real' model, we use
 for the inference a different one,  
 the isothermal sphere (chosen here in its singular, $\rho \sim r^{-2}$, version).

For the `real' model we take a `concentrated' sphere (CS) with a steeper density profile, 
$\rho \sim r^{-3.4}$, and an anisotropic velocity dispersion.  This density
profile (for $r\agt R_{\odot}$, the radius of the Sun's orbit) is commonly 
associated with the stellar halo (or `spheroid') that
consists of old, metal-poor stars.  While the local density of the luminous
halo is observed to be low $\sim 10^{-4}M_{\odot}/$pc$^3$ \cite{bss.83}, it is
possible that the stellar halo has a signifficantly more massive, dark (though
plausibly baryonic) counterpart of a similar structure. 
[A massive, dark `spheroid' of local density $\sim 10^{-3}M_{\odot}/$pc$^3$
was proposed in the past as a dynamically relevant component
of the galaxy \cite{calost.81,rk.88}. Microlensing by `spheroid' objects has been discussed
by, e.g., Giudice, Mollerach \& Roulet (1994) and De R\'{u}jula {\it et al.} (1995).]
Along with the $r^{-3.4}$ density profile, the radius-dependent velocity dispersion
anisotropy (see Section 2) adopted for our CS model describes 
rather accurately a well known stellar halo population, the  
blue horizontal branch field stars, BHBFS
\cite{jesper.obs}. The CS model reflects the possibility that,
just like the spatial distribution of the BHBFS, the massive objects'
 distribution may differ from that of the total (luminous + baryonic + nonbaryonic
 dark matter) halo mass.  
 
 As a result of our `experiment' we find that, 
 although $\alpha$ and $\beta$ are weakly affected 
 by the halo model ambiguity,
 the inferred $\bar{\mu}$ is on the average about 60\% greater
 than the real value.
 It is important that unless $N\agt 1000$, the statistics based 
 only on event durations does not allow us to distinguish between the two 
 massive objects' halo structures;
 the differences between them are submerged in the statistical noise.                                                 
 
 These results indicate
 that the uncertainty of the inferred average mass will be difficult
 to reduce below about the factor of 2.   A similar conclusion, with similar magnitudes
 of relative errors, holds for the massive object halo's local density $\rho_o$ and
 the total halo mass between the solar orbit and LMC, $M_{\rm tot}$.
 The unresolvable (at $N < 1000$) ambiguity is characteristic of a broad range of halo models
 that one might choose as the `real' ones instead of CS, although the ensuing uncertainty
 of the inferred average dark objects' mass and their total halo mass may be smaller than that
 obtained in the specific case of the CS/SIS ambiguity.        
 [Since the inferred halo density in the vicinity of the Sun for the CS
 model is about twice the value for the isothermal sphere, it may be possible
 to rule out 
 the more concentrated (e.g. $\rho\sim r^{-3.4}$ at $r\agt R_{\odot}$) 
 halo models on the basis of 
 dynamical arguments even at $N\alt 100$
 {\it if} the MHO halo indeed
 turns out to be very massive, i.e., approaching the total halo
 mass. For more detail see \cite{jd.2}.]
 These results are supported
 by our maximum likelihood simulations presented in Section 5, where the
 (appropriately parametrised) halo model is treated as unknown and its 
 parameters are varied together with the mass function parameters: only
 at  $N\agt 1000$ events do the errors in $\bar{\mu}$, $\rho_o$ and
 $M_{\rm tot}$ fall below 50\%.
 
 Given the current detection rate of a few events per year, and
 even hoping that it could be  increased in the future, the very 
 large number of events needed does not support optimism regarding an accurate
 inference of MHOs' masses and their halo fraction on the basis of
 LMC events only.  Probing the halo
 at various angles relative to the center of galaxy,
 e.g. through the observation
 of events toward  M31, could help
 distinguish between different halo models.  A considerable improvement
 might come from gaining more information from individual events
 by, e.g., parallax measurements (Refsdal 1966; Gould 1992, 1994, 1995b)
 that could put tight constraints on MHOs'
 spatial distribution and/or kinematics.
 
 This paper is organised as follows.  In Section 2 we review the
 halo models that will be used in the subsequent sections.  Section 3 outlines
 the derivation of microlensing rates.  In Section 4 we discuss the
 MHOs' mass function determination errors assuming a {\it known}
 halo model while in section 5 we study the effects of halo model
 uncertainties.   Finally, the Appendix provides some statements and formulae
 of mathematical statistics (along with the derivations) that are extensively
 used in the main body of the paper.

\section{Models of MHO distribution and kinematics}

In this paper we will consider a range of models  of the massive halo 
objects' distribution and velocities.  For simplicity, we will
restrict our discussion to spherically symmetric halos.
 One of the most commonly used
models is the isothermal sphere with the 
velocity dispersion constant throughout the halo the density profile
which is well approximated by
\begin{equation}
\label{isoth.core}
\rho (r) = \rho_o\frac{a^2 + R_\odot^2}{a^2 + r^2},
\end{equation}
where $a\approx 5\,$kpc is the `core' radius and $R_\odot = 8.5\,$kpc is
the distance of the Sun from the galactic centre.   Assuming that the total (luminous
+ dark matter) halo density is distributed according to expression 
(\ref{isoth.core}), one obtains the observed (approximately) flat rotation
curve for the galaxy.

The HMO mass distribution, however, need not follow that
of the total halo mass. One possibility is that the massive
objects may be more concentrated toward the galactic centre.
We may thus follow the clues provided by 
recent observations \cite{jesper.obs} of the blue horizontal
branch field stars (BHBFS) in the outer halo.  These observations
imply that the velocity dispersion changes from $\beta \equiv 
1 - \sigma_{\theta}^2 /\sigma_{r}^2 > 0$ at smaller distances
from the centre of the Galaxy to $\beta < 0$ at larger distances.
The radial velocity dispersion is well described
by the analytic fit
\begin{equation}
\label{disp.fit}
\sigma_{\rm r}^2 = \sigma_o^2 + \sigma_{+}^2 \left[\frac{1}{2} -
 \frac{1}{\pi}\tan^{-1}
 \left(\frac{r-r_o}{l}\right)\right],
\end{equation}
where the best agreement with the observations is achieved with
$\sigma_o = 80\, {\rm km}\,{\rm s}^{-1}$, $\sigma_{+} = 145\, 
 {\rm km}\,{\rm s}^{-1}$, $r_o = 10.5\,$kpc and $l = 5.5\,$kpc.
The BHBFS halo is close to spherical with the
density that is well modeled by the power law $\rho = \rho_o 
(R_{\odot}/r)^{\gamma}$, where $\gamma\approx 3.4$. 
\begin{figure}
\centering
\centerline{\epsfxsize= 8 cm \epsfbox[290 30 600 430]{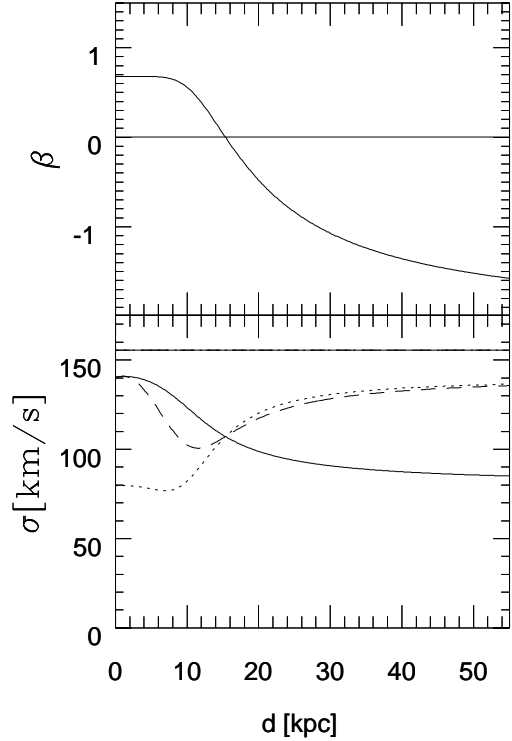}}
\caption{ Anisotropy parameter $\beta$ and velocity dispersion for 
 halo model CS as functions of the distance $d$ from the Earth in the
 direction of LMC; $\sigma_{\rm r}$ is given by the solid line, $\sigma_{\rm t}$
 by the dotted line and $\sigma_x$ [see paragraph preceding
 equation\ (\ref{new.gaussian})]
 by the dashed line.  The straight solid lines correspond to the
 singular isothermal sphere, SIS ($\beta = 0$, $\sigma = 156$ km/s) . 
 }
\end{figure}

From the Jeans' equation for spherical systems \cite{binney}
\begin{equation}
\label{jeans}
\frac{1}{\rho}\frac{d(\rho\sigma_{\rm r}^2)}{dr} +
    \frac{2\beta\sigma_{\rm r}^2}{r}
    = -\frac{d\Phi}{dr},
\end{equation}
where $\Phi(r)$ is the gravitational potential, we obtain
the tangential velocity dispersion
\begin{equation}
\label{disp.tan}
\sigma_{\rm t}^2 = \frac{1}{2}V_{\rm c}^2 - \left(\frac{\gamma}{2} -
   1\right)\sigma_{\rm r}^2 + \frac{r}{2}\frac{d\sigma_{\rm r}^2}{dr},
\end{equation}
where $V_{\rm c} = (-r d\Phi/dr)^{1/2}$ is the (roughly constant) rotation velocity.
Notice that given both the negative slope of the radial velocity
dispersion
\begin{equation}
\label{dsigdr}
r\frac{d\sigma_{\rm r}^2}{dr} = - \frac{1}{\pi}\frac{r}{l}
  \frac{\sigma_{+}^2}{1 + [(r -r_o)/l]^2},
\end{equation} 
and $\gamma > 2$, the tangential velocity dispersion will be
smaller than in the case of an isothermal sphere ($\alpha = 2$, 
$\sigma_{\rm r} =$ const. at large radii) with the same $V_{\rm c}$ 
[see Fig.\ 1].
This increased `pressure support' is merely a consequence of
the collisionless Boltzmann equation and --- in the final instance ---
the conservation of the phase-space volume (Liouville's theorem).

It is realistic enough --- and will prove quite convenient 
in the following --- to model the
velocity distribution by the Gaussian 
\begin{eqnarray}
\label{gaussian}
f(v_r, v_{\theta}, v_{\phi}) = \frac{1}{(2\pi)^{3/2}}
   \frac{1}{\sigma_{\rm r}\sigma_{\rm t}^2}\exp\left[-\frac{1}{2}\left(
   \frac{v_r^2}{\sigma_{\rm r}^2} +
   \frac{v_{\theta}^2 + v_{\phi}^2}{\sigma_{\rm t}^2}\right)\right],
    \hspace{-4cm} \nonumber \\ 
\end{eqnarray} 
where $\sigma_{\rm r}$ and $\sigma_{\rm t}$ are given by equations\
(\ref{disp.fit}) and\ (\ref{disp.tan}) for 
power-law density profiles.

In this paper we will adopt the following nomenclature for
 models of the massive objects' halo:
for the power-law halo with $\gamma =2$ and an isotropic
velocity dispersion we will use the familiar name, singular
isothermal sphere (`SIS'); if the density profile has a core (\ref{isoth.core}) and
the velocities are still distributed according to an isotropic version
of (\ref{gaussian}) with constant $\sigma_{\rm r} = 
\sigma_{\rm t} = V_{\rm c}/\sqrt{2}$, we will use the shorthand `CIS' 
(strictly, this 
model does not satisfy Jeans' equation);  the massive objects' halo
model corresponding to that of BHBFS with the power-law density profile
$\gamma = 3.4$ and the dispersion given by (\ref{disp.fit}) and Jeans'
equation will be called the `concentrated sphere' (`CS').

\section{Microlensing statistics}

The magnification of an LMC star due to the crossing of a
massive object near the line of sight from the earth
is given by\ \cite{paczynski.rev}
\begin{equation}
\label{magnif}
A(t) = \frac{u^2 +2}{u(u^2 + 4)^{1/2}},
\end{equation}
where 
\begin{equation}
\label{u}
u = \left(\frac{b^2 + v_{\rm n}^2 t^2}{R_{\rm E}^2}\right)^{1/2}.
\end{equation}
In the above equation $b$ is the impact parameter of the massive
object relative to the line of sight, $v_{\rm n}$ is object's
velocity orthogonal to the line of sight and 
\begin{equation}
\label{eins}
R_{\rm E} =
 \left[\frac{4 G m}{c^2} D x(1 -x)\right]^{1/2} =
 r_{\rm E} \sqrt{\mu x (1-x)}
\end{equation}
is the Einstein radius, where $D$ is the distance of the LMC
from the earth, $xD$ ($0 < x < 1$) is the object-earth distance,
$m = \mu M_{\odot}$ is object's mass and 
$r_{\rm E} = 3.2 \times 10^{9}\,{\rm km}$.

We define the duration of a microlensing event as 
$T = R_{\rm E}/v_{\rm n}$.  This is a {\it measurable}
quantity; it can be obtained as soon as one knows the maximum
magnification  $A_{\rm max} = A(u=u_{\rm min} = b/R_{\rm E})$ and, say, 
the time interval between half-magnifications $A = A_{\rm max}/2$
[see, e.g., \cite{derujula}].  The maximum magnification and
duration are the only two measurable quantities while
they depend on four parameters: $m, v_{\rm n}, x, b$.
Any information on properties of the massive object halo
can thus be  obtained only through a statistical analysis
of a sufficiently large number of events.

\begin{figure}
\centering
\centerline{\epsfxsize= 8 cm \epsfbox[-10 230 540 800]{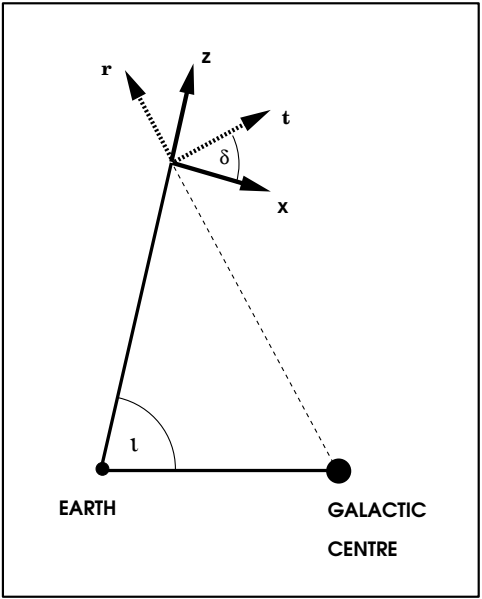}}
\caption{Coordinates and angles used in the derivation of the lensing
events rate.
}
\end{figure}

The statistical analysis proceeds from the
rate of microlensing events based on plausible
models of dark-mass halos, such as outlined in the last section.
Since the differential cross section per unit distance $x D$ for
a massive object to pass the line of sight between impact parameters
$R_{\rm E}u$ and $R_{\rm E}(u + du)$ is $2R_{\rm E}du$,
the rate in the more general case is given by 
\begin{eqnarray}
\label{rate}
\Gamma = N_{\ast}\int_0^1 D dx 
   \int d\mu \frac{dn}{d\mu}\;2R_{\rm E}\int_0^{u_{\rm th}} du_{\rm min}
 \int f_{\rm n}(v_{\rm n})v_{\rm n}dv_{\rm{n}}, \hspace{-2cm}
 \nonumber \\
\end{eqnarray}
where $N_{\ast}$ is the number of simultaneously observed stars,
$dn/d\mu$ is the differential number density of the massive objects,
$f_{\rm n}$ is the probability distribution of velocities
orthogonal to the line of sight and $u_{\rm th}$ is the threshold
for successful detection [the maximum amplification is required to
be greater than the threshold value, $A_{\rm max} (u_{\rm min}) > 
A_{\rm max} (u_{\rm th})$].  

In equation (\ref{rate}) we
ignore the motions of the earth and the source orthogonal
to the line of sight.  While enhancing the total event rate
by only a few percent in the case of sources in 
LMC, the observer's and
source's transversal motions lead to typically
shorter events \cite{griest.91}.  Although a proper
analysis of the results of {\it actual} microlensing searches
would  have to take this effect into account, we will
assume in this paper 
that the oberver and the sources are stationary. This
should suffice for an analysis that attempts
to isolate the effects of the halo objects'
mass function, spatial distribution and kinematics.
A more complete treatment would not change significantly
our main conclusions regarding the accuracy of 
the mass function inference; the considerably
greater computational effort does not seem
necessary at this point.

Both $f_{\rm n}(v_{\rm n})$ and $dn/d\mu$ depend on the chosen
model of the massive object halo.  
In order to describe the velocity distribution
at a point at distance $xD$ along the line of sight from the earth, 
we introduce a local coordinate system (see Fig.~2) so that the $x$ axis
is orthogonal to the line of sight and lies in the plane given by the earth, 
the center of the Galaxy and LMC, while the $z$ axis points toward 
LMC.  The distribution function (\ref{gaussian}) can now be rewritten 
in terms of the velocity components $v_x$, $v_y$ and $v_z$,
\begin{eqnarray}
\label{new.gaussian}
f(v_x, v_y, v_z) &=& \hspace{-0.2cm} \frac{1}{(2\pi)^{3/2}}
   \frac{1}{\sigma_{\rm r}\sigma_{\rm t}^2}
   \exp\bigg[-\frac{1}{2}\bigg(
   \frac{v_y^2}{\sigma_{\rm t}^2}  \nonumber \\
     \nonumber \\ &&\hspace{-1.1cm} +
   \frac{(\cos\delta\, v_x + \sin\delta\, v_z )^2}{\sigma_{\rm t}^2}
   + \frac{(\sin\delta\, v_x - \cos\delta\, v_z )^2}{\sigma_r^2}
   \bigg)\bigg],
    \nonumber \\
\end{eqnarray}
where $\sin\delta = (R_{\odot}/r) \sin\iota$ and $r^2 = R_{\odot}^2
+ (xD)^2 - 2 x D R_{\odot} \cos{\iota}$ ($\iota = 82^o$ for the LMC). 
The two-dimensional velocity
 distribution in the plane orthogonal to the line of sight is then
\begin{eqnarray}
\label{2D.gaussian}
f_{\rm t}(v_x , v_y) &\equiv& \int f(v_x, v_y, v_z) d v_z 
   \nonumber \\
  &=& \frac{1}{2\pi\sigma_{\rm x}\sigma_{\rm t} }
    \exp\left[-\frac{1}{2}\left(
   \frac{v_x^2}{\sigma_x^2} +
   \frac{v_y^2}{\sigma_{\rm t}^2}\right)\right],
\end{eqnarray}
where $\sigma_x^2 = \cos^2\delta\, \sigma_{\rm t}^2 + \sin^2\delta\,
  \sigma_r^2$.  It is now straightforward to obtain
  the distribution for velocity $v_{\rm n} = |\cos\phi\, v_x + \sin\phi\,
   v_y |$ orthogonal to the line of sight
\begin{eqnarray}
\label{fvn}
f_{\rm n}(v_{\rm n}) \hspace{-0.15cm}&=& \hspace{-0.15cm}
   \int_0^{2\pi}f_{\rm t}(v_{\rm n}\cos\phi,\,
 v_{\rm n}\sin\phi )\, v_{\rm n}\, d\phi 
   \nonumber \\
  &=& \hspace{-0.2cm}\frac{1}{2\pi\sigma_{\rm x}\sigma_{\rm t}}
     \int_0^{2\pi}d\phi\; v_{\rm n}e^{
     -\frac{v_{\rm n}^2}{4}\left[\frac{1}{\sigma_x^2}
     + \frac{1}{\sigma_{\rm t}^2} + \left(\frac{1}{\sigma_x^2}
     - \frac{1}{\sigma_{\rm t}^2}\right)\cos 2\phi\right]}
       \nonumber \\
  &=&  \frac{1}{\sigma_{\rm x}\sigma_{\rm t}} v_{\rm n}
        I_{o}\left(\left|\frac{1}{\sigma_x^2} - \frac{1}{\sigma_{\rm t}^2}
	\right|\frac{v_{\rm n}^2}{4}\right)
	e^{-\frac{1}{4}\left(\frac{1}{\sigma_x^2} +
	  \frac{1}{\sigma_{\rm t}^2}\right) v_{\rm n}^2},
	  \nonumber \\
 \end{eqnarray}
 where we have used the identity for the Bessel function of
 the zero'th order $2\pi I_{o}(x) =
   \int_0^{2\pi}\exp\,(x\cos\phi)\, d\phi$ \cite{abram}.
Of course, for an isotropic velocity dispersion one recovers
the familiar Maxwell distribution.

As we indicated in the last section, the massive object 
number density 
will be modeled either as the power law
\begin{equation}
\label{density}
n = n_{o} \left(\frac{R_{\odot}}{r}\right)^\gamma = n_o H(x),
\end{equation}
where
\begin{equation}
\label{Hx}
H(x) = \left[ 1 + x^2 \left(\frac{D}{R_{\odot}}\right)^2 -
  2 x \frac{D}{R_{\odot}} \cos\iota \right]^{-\gamma/2},
\end{equation}
or as the modification of (\ref{Hx}) for $\gamma = 2$ that includes a core 
(\ref{isoth.core}).  Throughout this paper we will assume that the
mass function $dn_o/d\mu $ is independent of the position in
the halo.

Recalling that velocity is related to the duration of a microlensing
event by $v_{\rm n} = R_{\rm E}/T = r_{\rm E}\sqrt{\mu x (1-x)}/T$,
we can change the variable of integration in (\ref{rate})
from $v_{\rm n}$ to $T$ and thus obtain the rate of event {\it detection}
\begin{equation}
\label{rate.T}
\Gamma = 2 N_{\ast} D r_{\rm E}^3 u_{\rm th}\int d\mu \frac{d n_o}{d\mu}
  \int dT \,\varepsilon (T) F(\mu/T^2),
\end{equation}
where
\begin{eqnarray}
\label{PmuT}
F(\mu/T^2) &=& \left(\frac{\mu}{T^2}\right)^{3/2}
                    \int_0^1 
                     [x(1-x)]^{3/2}
  H(x) \nonumber \\
  && \hspace{1.8cm} \times\; 
        f_n \left[r_{\rm E}\sqrt{x(1-x) \mu/T^2}\right] dx.
   \nonumber \\
\end{eqnarray}
In the last expression we have introduced the detection efficiency
$\varepsilon (T)$, i.e. the fraction of all events of duration $T$
and satisfying $u < u_{\rm th}$
that will be detected with the available techniques.  In this paper
we will use the following approximate expression for the
detection efficiency
\begin{eqnarray}
\label{macho.eff}
\varepsilon (T) = \left\{\begin{array}{ll}
     0.3\; e^{-(\ln(T/T_{\rm m}))^{2.6}/2.54}, & \mbox{$T > T_{\rm m}$}
     \\  
     \\
     0.3\; e^{-|\ln(T/T_{\rm m})|^{1.9}/3.56}, & \mbox{$T < T_{\rm m}$}
     \end{array}
     \right.
\end{eqnarray}
where $T_{\rm m} =$ 75 days.  This expression is in  good numerical agreement
with the efficiency quoted by the MACHO team\ \cite{alcock.96}
for their two-year LMC microlensing detection results.

\begin{figure*}
\centering
\centerline{\epsfxsize= 17 cm \epsfbox[0 120 640 680]{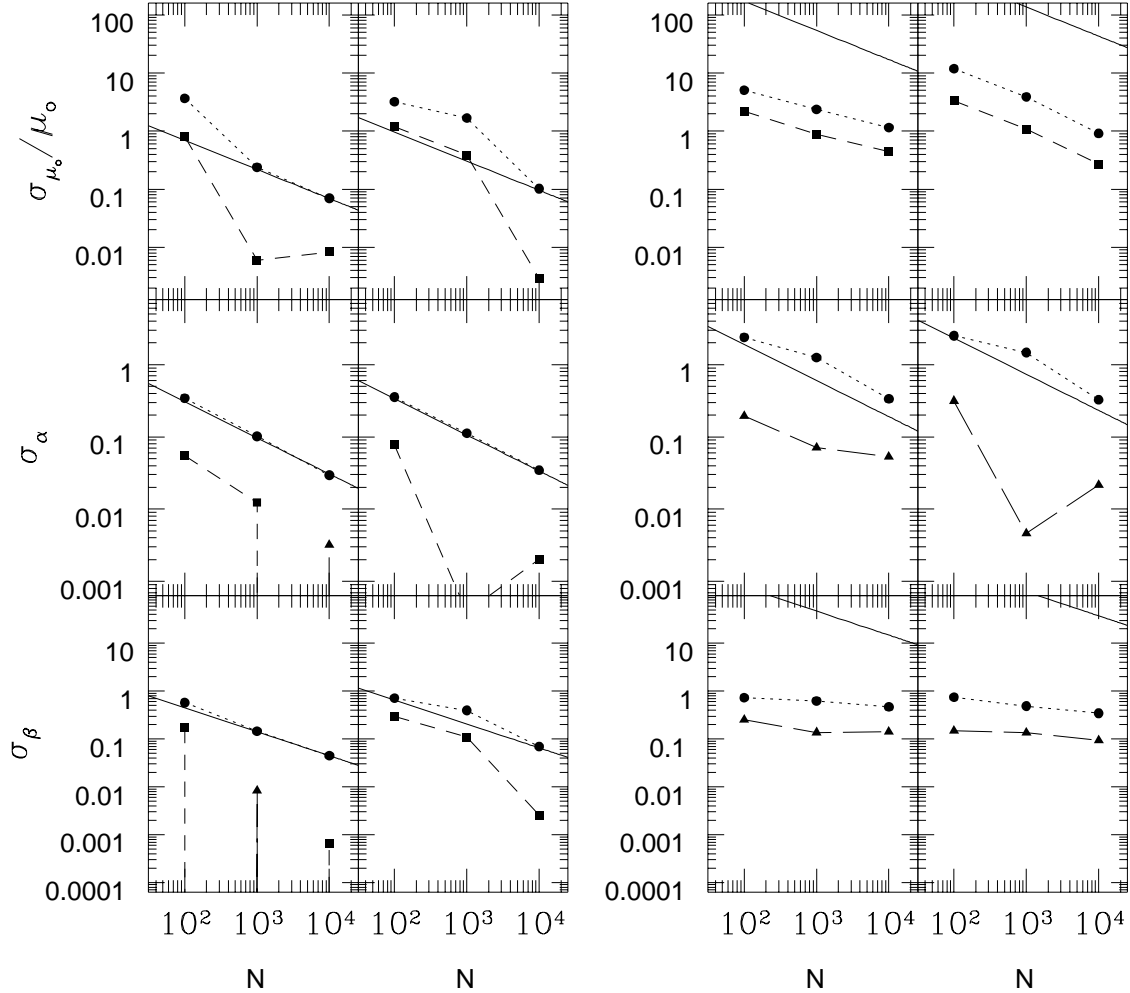}}
\caption{ Dependence of variances $\sigma_{\mu_o}$, $\sigma_{\alpha}$
     and $\sigma_{\beta}$ on the number of detected events for
     $\mu_o = 0.86$, $\alpha = -2$ and $\beta = 2$ (left panel) and
     $\mu_o = 0.06$, $\alpha = 1$ and $\beta = 2$ (right panel). [In both
     cases the average mass $\bar{\mu}=0.4$.]
     In each panel the left column is based on a flat ($\varepsilon =$const.)
     detection efficiency, while the right-hand column assumes
     the MACHO-type efficiency (\ref{macho.eff}).
     The solid circles connected by dotted lines represent the variances
     as obtained by Monte-Carlo simulations.  The mean shift of the
     Monte-Carlo inferred values relative to the `real' parameters (bias) is
     given by squares connected by short-dashed lines (positive bias)
     or triangles connected by long-dashed lines (absolute value of
     negative bias).
     The variances in
     the Cramer limit are given by solid lines.
   }
\end{figure*}

As a model for the massive objects' mass function we choose
a simple power law [see e.g. \cite{mao.pacz}]
\begin{equation}
\label{massfun}
p_{\mu}(\mu ) d\mu = \frac{1}{C_{\beta}(\alpha)} 
          \frac{\mu^{\alpha}}{\mu_o^{\alpha +1}} d\mu,
\end{equation}
for the probability that the mass of a star lies
in the interval $M_{\odot}d\mu$.  
In the simplest possible case the mass function is specified, apart 
from the exponent $\alpha$,  by the range of masses 
$\beta\equiv \log_{10}(\mu_{\rm max}/\mu_{\rm min})$, where
$\mu_{\rm max}$ and $\mu_{\rm min}$ are the upper and lower limits
of the range, and by the geometric mean
$\mu_o = (\mu_{\rm max}\mu_{\rm min})^{1/2}$.  The normalisation
constant in equation (\ref{massfun}) is then given by
\begin{eqnarray}
\label{cbetalp}
C_{\beta}(\alpha) = \left\{\begin{array}{ll}
            \beta\ln 10 & \mbox{$\hspace{0.5cm}\alpha =-1$},\hspace{-1cm} 
	    \nonumber \\
	    \nonumber \\
	    \frac{1}{\alpha +1}\left[10^{\beta(\alpha +1)/2}
	      - 10^{-\beta(\alpha +1)/2}\right]
	      & \mbox{$\hspace{0.5cm}\alpha \neq -1$},\hspace{-1cm}
	                     \nonumber
	         \end{array}
		 \right. \nonumber \\ 
\end{eqnarray}
while the massive objects' mass density near the Sun is related
to their number density by $\rho_o = n_o\mu_o M_{\odot} C_{\beta}
(\alpha + 1) /C_{\beta}(\alpha)$.

\section{Determining MHO mass function: fixed halo model}

Given a sufficient number of detected microlensing
events, one can attempt to infer the mass function 
of the lensing massive objects. In this section we
will estimate the accuracy of such an inference.
In different words, we will try to estimate just
what is the `sufficient' number of events for
a reliable mass function determination. 

In this section we make the important assumption that
the halo structure, given by MHO's density profile
and kinematics, is known (the consequences of the
halo uncertainty will be discussed in the next section).  We then
simulate the 
inference of the mass function parameters
$\mu_o$, $\alpha$ and $\beta$ from samples of a fixed
number $N$ of microlensing events of durations $T_i$, $i=1,N$.
We use the maximum likelihood method [for an alternative method of
mass function inference,
based on mass momenta see \cite{derujula,jetzer94,jetzer96}], where we
maximise the function 
\begin{equation}
\label{lik}
l(\{T_i\}| \mu_o, \alpha, \beta) = \Sigma_{i=0}^{n}
     \ln P(T_i | \mu_o, \alpha, \beta)
\end{equation}
with respect to the parameters $\mu_o$, $\alpha$ and $\beta$
[see \cite{gould.ML} for a different yet equivalent formulation of
the method];
we denote the values of parameters at the maximum of $l$
by $\hat{\mu}_o$, $\hat{\alpha}$ and $\hat{\beta}$.
In equation (\ref{lik}) $P(T | \mu_o, \alpha, \beta)$ is
the normalised ($\int P(T |\cdots )\, dT =1$) differential
event rate, $P(T)\propto d\Gamma /dT $.  

Given the power-law model of the MHO mass function (\ref{cbetalp}),
the rate $\Gamma$ [see equation (\ref{rate.T})] can be rewritten as
\begin{eqnarray}
\label{rate.y}
\Gamma &=& 2 N_{\ast}D r_{\rm E}^3 u_{\rm th} n_o \int dT \nonumber \\
 	& & \hspace{1cm}\times
	     \frac{\varepsilon (T)T^{2(\alpha +1)}}
	            {C_{\beta}(\alpha) \mu_o^{\alpha +1}}
    \int_{\mu_o \frac{10^{-\beta/2}}{T^2}}^
          {\mu_o \frac{10^{\beta/2}}{T^2}} y^{\alpha} F(y) dy,
\end{eqnarray}
where $y \equiv \mu/T^2$.

In both panels of Fig.\ 3 we have assumed that the MHO 
distribution and kinematics are well described by the isothermal sphere
with a core [CIS, see equations\ (\ref{isoth.core}) and (\ref{gaussian})]
and that we use just this (the `real') halo model for
the MHO mass function inference.
The underlying 
(`real') mass function in the left-hand panel is given by
 $\mu_o = 0.86$, $\alpha = -2$ and 
$\beta = 2$, while that in the right-hand panel corresponds to
$\mu_o = 0.06$, $\alpha = 1$ and $\beta = 2$. (The average mass in
both cases is $\bar{\mu}=0.4$.)  The left columns
of each of the two panels show results for $\varepsilon (T) =$ const.,
while in the right-hand side columns the detection efficiency is assumed to
be of the form\ (\ref{macho.eff}).

In the figure we compare results of Monte-Carlo simulations 
(root-mean-square variations from the mean values of the
inferred parameters are given by solid circles connected by dotted lines)
for $N=$ 100, 1000 and 10000   
with the error estimates obtained in the so called Cramer limit
(solid lines). The Cramer limit 
error $\sigma^{\rm c}(c_{\mu})$  in the determination of
parameter $c_{\mu}$ from $N$ data points is given by
\begin{equation}
\label{cramer.err}
\left[\sigma^{\rm c}(c_{\mu})\right]^2 = 
   \left[ I^{(N)}_{\mu\mu}({\bf c}) \right]^{-1},
\end{equation}
where the expression on the right-hand side is the $\mu\mu^{'{\rm th}}$
component of the inverse of the information matrix [see Appendix, 
equation\ (\ref{I.def})]
\begin{eqnarray}
\label{info.real}
I^{(N)}_{\mu\nu}({\bf c}) &=& N\int \left( \frac{\partial}{\partial c_{\mu}}
              \ln P(T|{\bf c})\right)
    \left( \frac{\partial}{\partial c_{\nu}}\ln P(T|{\bf c})\right) 
    \nonumber \\ \nonumber \\ 
   & & \hspace{2cm}\times\; P(T|{\bf c}) \; dT,
\end{eqnarray}
where ${\bf c}$ denotes the three parameters $\mu_o$, $\alpha$ and $\beta$.

For the negative slope ($\alpha =-2$)  
the errors in $\alpha$ and $\beta$
are rather small at $N > 100$ and, not surprisingly,
the Cramer limit is in good agreement with the results of Monte-Carlo
simulations.  As for $\mu_o$, a respectable accuracy can be
achieved only for $N\agt 1000$ if $\varepsilon$ is flat or for
even a larger number of events
if the detection efficiency is of the MACHO type.  Notice,
however, that convolution with a MACHO-type $\varepsilon(T)$ typically results 
in a relatively moderate error increase at a fixed number $N$ of detected events. 
 Still, the parameter estimation
would take considerably more observation time due to the simple fact that this
detection efficiency allows us to detect only 1/4 to 1/3 of
all microlensing events, given the mass range discussed in this paper.

\begin{figure}
\centering
\centerline{\epsfxsize= 9 cm \epsfbox[40 400 600 780]{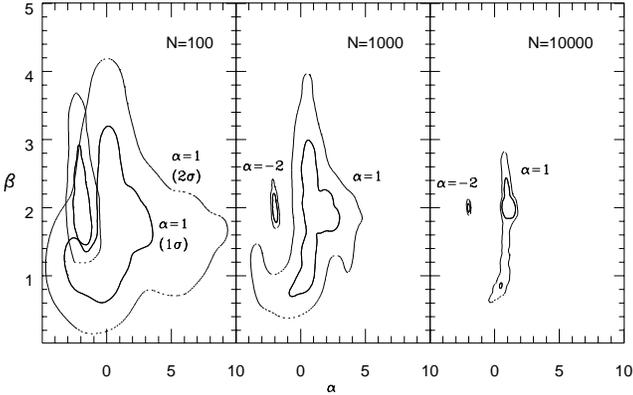}}
\caption{ `$1\sigma$' (68\% confidence level) and `$2\sigma$' (95\% confidence)
 contours obtained by Monte-Carlo simulations of $\alpha$ and $\beta$ inference
 for both $\alpha = -2$ 
and $\alpha =1$.  The third parameter, $\mu_o$, has been `integrated ' over.  
The panels show results for (left to right)
100, 1000 and 10000 events.  In all cases $\beta = 2$, $\bar{\mu} = 0.4$
and $\varepsilon(T) =$ const. Notice the smaller size and more rapid
shrinking with increasing $N$ in the case of the contours centered on $\alpha =-2$,
$\beta = 2$.  At $N=1000$ and especially at $N=10000$ the contours are tightly
concentrated around the `true' value $\alpha =-2$.  
   }
\end{figure}

The errors are much larger in the positive slope ($\alpha = 1$) case.
The determination of the slope itself is rather inaccurate, 
$\sigma_{\alpha}/|\alpha | > 1$, unless $N$ approaches 10000 (see the
Monte-Carlo results). Although only moderately 
large ($\sigma_{\beta}\approx 1$)
at small numbers of
events ($N\approx 100$), the uncertainty of $\beta$ decreases 
very slowly for larger $N$.  In addition,
the statistical error in $\mu_o$ is hopelessly large even
for the largest number of events considered in Fig.\ 3  (notice
that $\mu_o = 0.06$).

The rather grim prospects for the inference of the $\alpha = 1$
mass function are further illustrated by the contour plots
of Fig.\ 4.  Closely related to the large errors, in about 30\% of the
Monte-Carlo simulations at 100 events, the maximum likelihood 
procedure results in a delta-function best fit for the MHO mass 
distribution.  [This fraction falls to 2-3\% at $N$=1000, and then
to $0$\% at $N$=10000.]  This is indicated by either the
inferred $\alpha$ tending to $+\infty$ or (much less frequently)
$\beta$ approaching 0.  In other words,
the statistical errors are so large as to have a good chance of
concentrating events' duration scatter to what may look like
an effect of a single-mass MHO population.  These cases were
excluded when the statistics of Monte-Carlo results was computed.

The irregular shape and great extent of the contours 
explains why Cramer limit is such a poor approximation in this
case. Indeed, the Cramer limit depends only on the properties
(more specifically, the first derivatives with respect to the
parameters ${\bf c}$) of the distribution function in the immediate
vicinity of the `real' parameters.  In the $\alpha = 1$ case
the Cramer limit gives large errors indicating that the function
is very insensitive to small changes in the parameters.
However, once we shift away from the original parameters, the
nonliner distorsions of the distribution function are probed
and this may result in {\it smaller} deviations of the inferred parameters
then one would expect on the basis of the Cramer limit estimates.  
[This is indeed the
case for $\mu_o$ and $\beta$ at $\alpha = 1$ 
(see again Fig.\ 3; the Cramer limit is given by
the solid lines).]  Since the distribution function $P(T|{\bf c})$ is
intrinsically so little dependent on small changes in the parameters 
${\bf c}$, convolving with a MACHO-type detection efficiency $\varepsilon (T)$
will not change the estimation accuracy significantly, as can also
be concluded from Fig.\ 3.

\begin{figure}
\centering
\centerline{\epsfxsize= 9 cm \epsfbox[40 57 600 760]{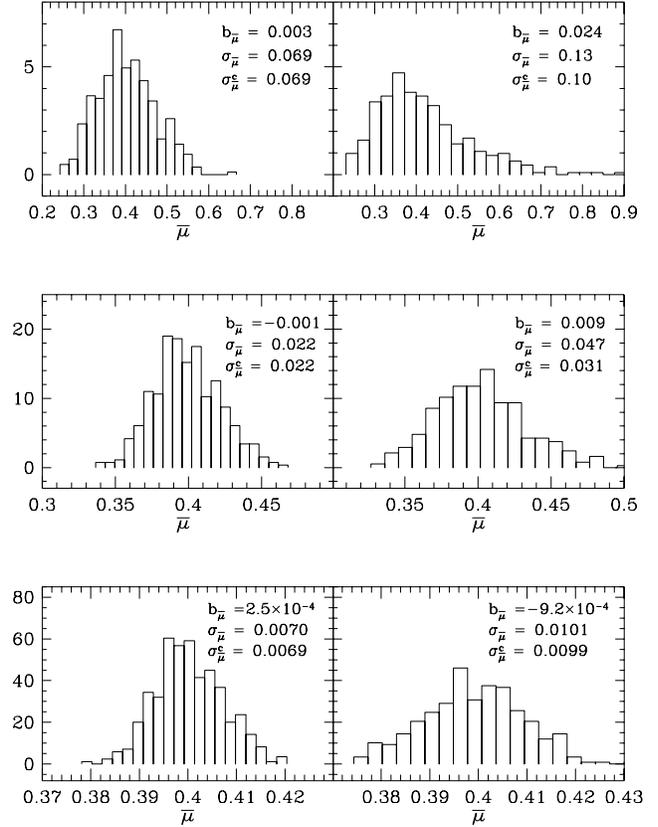}}
\caption{ Histograms of inferred average masses $\bar{\mu}$
     obtained  by Monte Carlo simulations for the  
     `real' parameters
     $\mu_o = 0.86$, $\alpha = -2$ and $\beta = 2$ (thus $\bar{\mu}
     = 0.4$) in the
     case of uniform sensitivity (left) and the MACHO-type sensitivity
     [equation (\ref{macho.eff})] (right) for $N=100$ (top), 1000
     (middle) and 10000 (bottom) events.  The numbers in the upper right
     corners indicate mean shifts ($b$) of the inferred parameters relative
     to the `real' values, root-mean-square variances ($\sigma$)
     with respect to the means and the Cramer-limit errors 
     ($\sigma^{\rm c}$).  The underlying halo model
     is the isothermal sphere with a core (CIS).
   }
\end{figure}

The very large errors in the estimation of $\mu_o$ both for
$\alpha=-2$ and (especially) $\alpha=1$ prompt us to seek
a combination of the three parameters $\mu_o$, $\alpha$ and $\beta$ that
can be inferred with greater accuracy.  Not surprisingly, the average
mass $\bar{\mu}$ (the first moment of the MHO mass distribution) is
just such a quantity.  In Fig.\ 5 we plot the histograms of
inferred $\bar{\mu}$ for the negative slope ($\alpha = -2$) case 
obtained by the same Monte-Carlo simulations (400 simulations
for each set of parameters, as is sufficient to obtain a relatively smooth distribution
of points in the $\alpha$ -$\beta$ plane) 
whose results were shown in Figs.\ 3 and 4.
We can see that even for $N=100$ events the average mass can be inferred
with decent accuracy.  Notice that the Cramer-limit errors are 
in good agreement with the Monte-Carlo results.  The same conclusion ---
as well as similar values of the errors --- holds for the positive slope 
$\alpha = 1$.
  
\begin{figure}
\centering
\centerline{\epsfxsize= 9 cm \epsfbox[50 40 360 512]{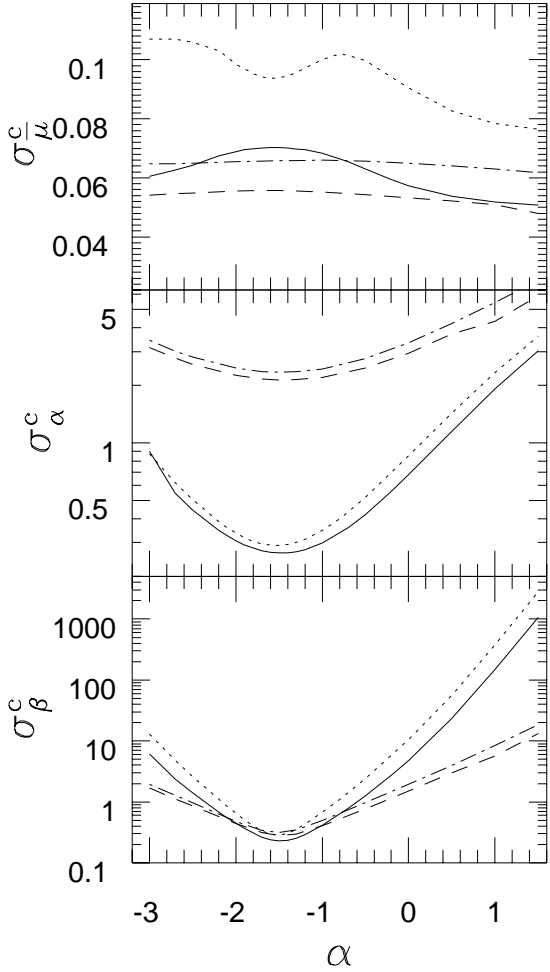}}
\caption{ The dependence of the Cramer-limit errors on $\alpha$.  The errors
(normalised to 100 events)
 for a `broad' range, $\beta =2$, mass function are shown as the solid
 [$\varepsilon (T) =$ const] and dotted [MACHO-type detection
 efficiency, $\varepsilon (T)\neq $const.]
  lines.  For a `narrow' mass range, $\beta = 1$,
  the errors are shown as the
 dashed and dot-dashed lines for the respective detection efficiences.
   }
\end{figure}

As we could see in Fig.\ 3, Cramer limit gives a poor estimate of the actual
maximum-likelihood errors when those errors are large.  However,
if in some regions of the parameter space we detect peculiarly large
Cramer-limit errors, that implies (as we have seen above) a low
 sensitivity of the distribution function $P(T|{\bf c})$ to small
shifts in the underlying parameters and correspondingly indicates
larger maximum-likelihood errors.  By contrast, if
the Cramer-limit errors are small
they can indeed be used as realistic estimates of the actual inference
uncertainties.  

In Fig.\ 6 we show Cramer-limit errors, normalised
to 100 events as functions of the `real' $\alpha$.  The `real' $\beta$
is chosen to be 2 (`broad' MHO mass function) or 1.  Apart from
the larger errors for the MACHO-type $\varepsilon(T)$ compared to
the flat $\varepsilon$ errors, another easily predictable feature is the
much poorer accuracy of the slope $\alpha$ determination for a narrow-range 
($\beta =1$) mass function in comparison with the broad-range function: the slope
is sampled much better in the latter case.

Somewhat less obvious is that the accuracy of shape
(slope $\alpha$ and range $\beta$) determination should peak as sharply 
at $\alpha = -1.5$ as 
we observe in Fig.\ 6.  Indeed, one would expect large uncertainties in
the slope and range determination if the slope $\alpha$ is of large magnitude
--- positive or negative --- thus bringing the mass distribution close to the
delta function limit. The fact that the transition region between 
the two extremes
is centered  on $\alpha=-1.5$ can be understood as follows.
 
For large values of $\beta$ expression 
$T P(T)$ as a function of of $\ln T$
 has a broad plateau [$d (T P(T))/d\ln T\approx 0$]  for $\alpha =
- 3/2$ as can be
seen from equation\ (\ref{rate.y}): since $F(y)\sim y^2$ for 
$y\ll (\sigma/r_{\rm E})^2/\bar{\mu} $
and $F(y)\sim 1/y$ for $y\gg (\sigma/r_{\rm E})^2/\bar{\mu}$ ($\sigma$ is
typical velocity dispersion), the last integral depends relatively 
weakly on $T$ for
$10^{-\beta/4}\bar{\mu}^{1/2}r_{\rm E}/\sigma < T <
10^{\beta/4}\bar{\mu}^{1/2}r_{\rm E}/\sigma$.  
\begin{figure}
\centering
\centerline{\epsfxsize= 9 cm \epsfbox[10 30 590 760]{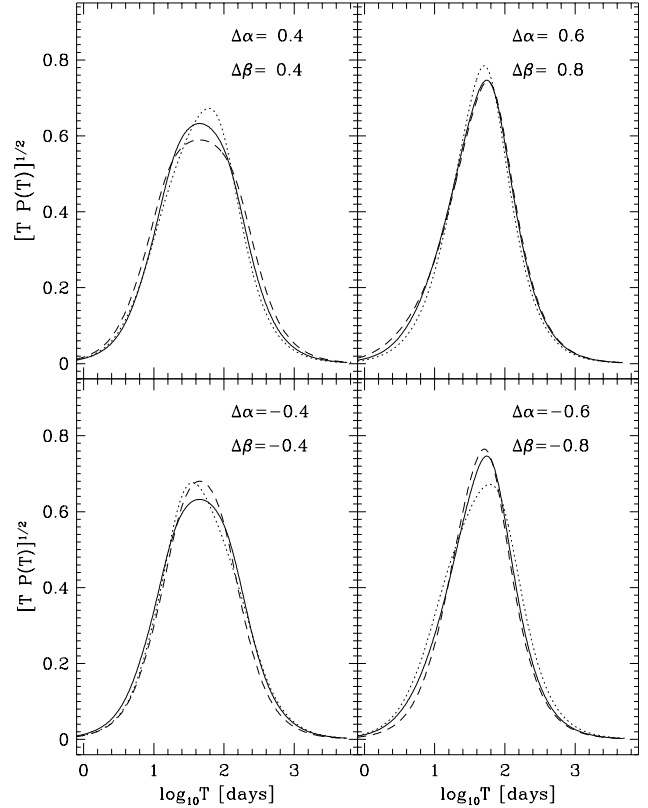}}
\caption{Changes in curves $[TP(T)]^{1/2}$ resulting from small shifts in
$\alpha$ (dotted) and $\beta$ (dashed) for $\alpha=-3/2$ (left)
and $\beta=-1/2$
(right).  The values of parameter shifts are indicated in the figure.
The reference curves (solid lines) are chosen to have $\beta=2$. Notice that we
had to choose larger magnitudes of parameter shifts in order to
produce discernible changes for $\alpha = -1/2$. $\varepsilon(T) =$ const.
is assumed.
}
\end{figure}
Since the components of the information matrix (\ref{info.real}) can be viewed
as scalar products of  the derivatives $\left(\sqrt{TP(T)}\right)_{,\mu}$
(where $\ln T$ is used as the measure), larger distorsions
of $\sqrt{TP(T)}$ resulting from small changes in the underlying
parameters will lead to smaller errors of parameter determination. 
This is illustrated in Fig. 7 where we show the 
curves $[TP(T)]^{1/2}$ for $\beta$ in the vicinity
of $2$ and mass function
slope  near $\alpha=-3/2$ (upper and lower left) and $\alpha = -1/2$
(upper and lower right).  The dotted ($\Delta\beta =0$) and dashed ($\Delta\alpha
= 0$) curves show the distorsions resulting from shifts in the parameters
as indicated in the figure.  Although $\beta$ is not large, and
accordingly, the `plateau' is not very broad, one can clearly recognise the effects 
of $\alpha$ being equal to $-3/2$:
it is due to the relatively large spread in $\ln T$ (plateau) that the 
curve is so sensitive to small parameter changes.  In addition, the consequences
of the shifts in $\alpha$ and $\beta$ can be clearly distinguished.  Thus
for $\Delta\alpha = 0.3$, the mode of the curve shifts towards longer times;
the rough symmetry of $\sqrt{TP(T)}$ at the
plateau is broken and now longer events dominate. By contrast, the change
in $\beta$ only supresses ($\Delta\beta > 0$) or sharpens ($\Delta\beta < 0$)
the curve without shifting its mode. On the other hand, for $\alpha=-1/2$
a positive shift in $\alpha$ is closer to producing 
effect similar to that of a negative $\Delta\beta$.  This degeneracy
gives rise to a small value of the determinant of the 
information matrix $I_{\mu\nu}$ which
enhances the errors in $\alpha$ and $\beta$.

The flatness of $\sqrt{TP(T)}$ in the vicinity of $\alpha = -3/2$ corresponds
to roughly equal contribution of the MHO's with masses $\mu > \mu_o$
and the ones with $\mu < \mu_o$ [see also \cite{mao.pacz}] to the
total event rate.  As we move toward larger $\alpha$, the rate is dominated
by larger masses (or, if we reduce $\alpha$, by smaller masses).  
This simple fact
helps explain an interesting feature of the bottom panel of Fig. 6:
while  at $\alpha=-3/2$ the error in 
$\beta$ is slightly lower for
$\beta =2$ than for $\beta=1$ (see argument above: the plateau is broader
for $\beta = 2$), this error increases more rapidly for $\beta = 2$
as we shift away from $\alpha = -3/2$. This is a consequence
of the `tail' (in terms of contribution to the event rate)
of the mass function being less accessible
[i.e. more distant from the (effective) concentration of masses at the opposite
end] in the case of the larger $\beta$.    


\section{Effects of the uncertainty of MHOs' 
           spatial distribution and kinematics}

In this section we will study the extent to
which the inference of the MHO mass function will be hampered
by the unavoidable uncertainty of the massive objects' halo
structure, i.e. their distribution and kinematics.
For simplicity, we will consider only spherically symmetric halos.

At first, we will estimate  how much the
inferred parameters of the mass function could  be offset {\it systematically}
relative
to the real values if instead of fitting the event duration distribution
function $P_o(T)$ based on the `true' halo, we use the distribution
function $P(T)$ based on a `false' halo model. 
As we explain in the Appendix, a reasonable estimate of this shift
in the parameters can be obtained by finding the mass function
parameters ${\bf c}$ for which the expression
\begin{equation}
\label{Psi.bias}
\Psi({\bf c}|{\bf c}_o) \equiv \int P_o (T | {\bf c}_o) 
          \ln P( T |{\bf c})\, d T
\end{equation}
is maximised (${\bf c}_o$ denotes the `real' mass function parameters).

\begin{figure}
\centering
\centerline{\epsfxsize= 8 cm \epsfbox[50 70 480 640]{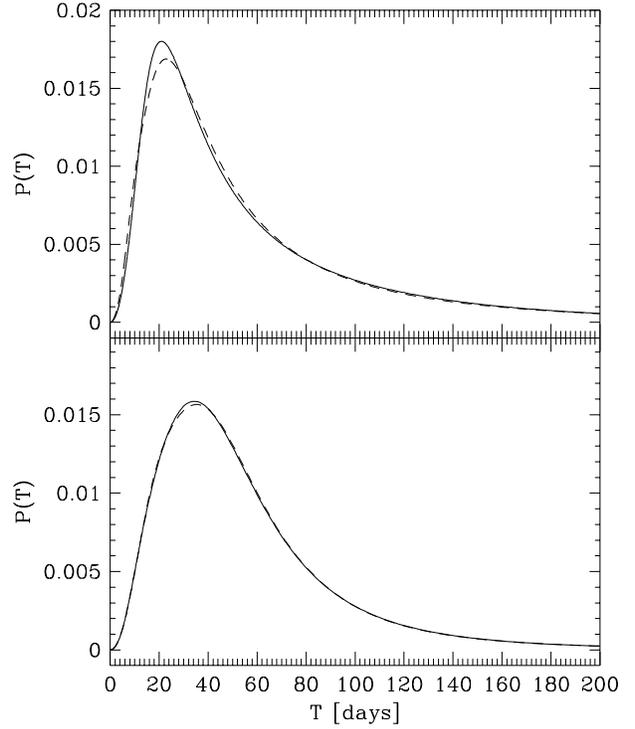}}
\caption{The differential event rate (upper panel) for $\bar{\mu}_o=0.4$, 
$\alpha_o=-2$, $\beta_o =2$
and the `real' halo given by the `CS' model (dashed line) and its closest SIS
 match ($\bar{\mu}=0.630$, $\alpha = -1.874$, $\beta =2.035$) [solid line]
 obtained by maximising\ (\ref{Psi.bias}) .   In the lower panel,
 $\bar{\mu}_o=0.4$, $\alpha_o= 1$, $\beta_o =2$, while the SIS-based best fit
 is $\bar{\mu}=0.624$, $\alpha =-0.188$ and, $\beta =1.415$. 
 The two curves in the latter case are virtually indistinguishable.
} 
\end{figure}

In Fig. 8 we show distributions
$P_o(T)$  based on the `concentrated
sphere' (CS) halo model {\it along with} their best matches from among the distributions
$P(T)$ based on the SIS model for parameters indicated in the figure caption.
[Here as throughout this section we use $\varepsilon (T) =$ const.]  
While the two curves in the upper panel are  close to each other,
the ones in the lower panel are virtually indistinguishable.
This is a typical situation for a broad range of the (`real')
mass function parameters: by shifting the mass function parameters we can 
mask to a great extent the effects of changes in the halo model.  Conversely, as long
as we lack {\it independent} information about the MHOs' distribution
and kinematics, there always will be significant errors in 
the determination of the mass function from event durations only.
The relatively less accurate match of the two curves in
the upper panel  is characteristic of the values of $\alpha$
that  allow an accurate inference of the mass function according to the
results of the last section.
We will discuss below in this section how we can take advantage of this 
residual difference to obtain a more accurate measure of the mass function.
At present we will assume that we are indeed in error regarding
the MHO distribution and kinematics and are thus  using a false (SIS)
halo model instead of the right (CS) one.  
 What are the magnitudes
of the systematic errors that ensue?

In Fig. 9 we can observe the  dependence of the systematic shift, described
above, on the value of $\alpha_o$ for $\beta_o =1$ (solid curve) and
$\beta_o = 2$ (dotted curve).  These results were obtained by maximising
expression (\ref{Psi.bias}) for $\varepsilon(T)=$
const.

\begin{figure}
\centering
\centerline{\epsfxsize= 8 cm \epsfbox[280 40 600 610]{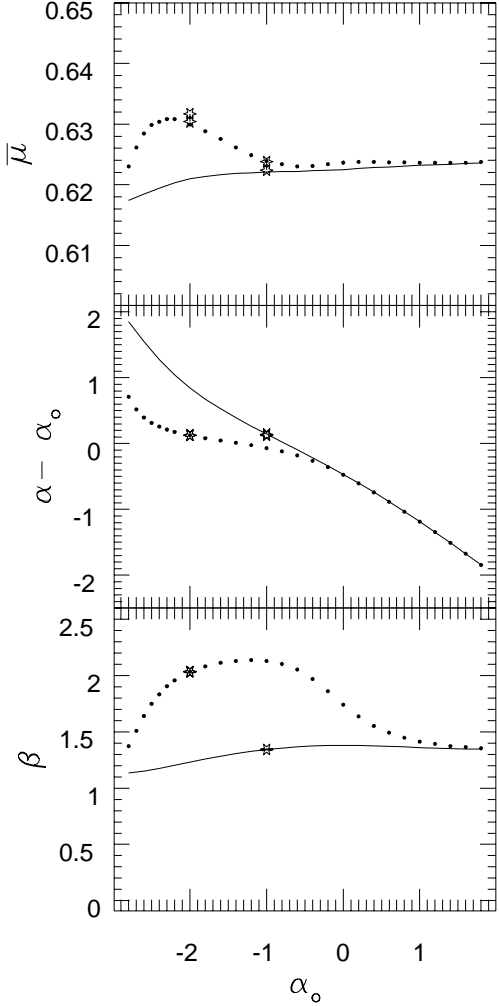}}
\caption{Systematic shift (bias) in the inferred parameters relative to
their `real' values as a function of the real $\alpha_o$.  
It is assumed that the real halo is of the `CS' type,
while the singular isothermal sphere (SIS) is used in the inference of $\bar{\mu}$,
$\alpha$ and $\beta$.  For the dotted lines $\beta_o =2$, while for the 
solid lines $\beta_o = 1$.  In all cases $(\bar{\mu})_o = 0.4$.  The curves
were obtained by maximisation of expression\ (\ref{Psi.bias}).
The stars indicate the results of Monte-Carlo simulations.  The vertical
segments bordered by stars in the plot of $\bar{\mu}$ give the 
$1\sigma$ error for the Monte-Carlo simulations.
} 
\end{figure}

We first notice the upward shift in the inferred average mass $\bar{\mu}$ by
about 60\%.  The inferred mass scales as $\mu\sim \sigma^2/ 
\langle x(1-x)\rangle$,
where $\sigma$ is the typical velocity dispersion in the halo
and $\langle x(1-x)\rangle$ denotes the average of the quantity over the
differential event rate.  By using the SIS instead of the `right' CS halo
model, we gain a factor of $2-3$ due to the increased $\sigma$ (see Fig. 1),
while the mass estimate is reduced due to the larger $\langle x(1-x)\rangle$
(the SIS halo is more extended).  The net result is a shift in $\bar{\mu}$
that depends only weakly on $\alpha_o$ and $\beta_o$. [The segments bordered by
stars show results of Monte-Carlo simulations (10000 events)
up to their statistical errors
obtained from 400 simulations.
The agreement with the results obtained by minimising (\ref{Psi.bias})
is good but not perfect.  See Appendix for a brief discussion of the
limitations of the method based on 
equation (\ref{Psi.bias}).]

We also notice that for $-2 \alt \alpha_o \alt 0$ both $\alpha$ and $\beta$ are
relatively weakly affected by the uncertainties of the halo model.  This may
be expected on the basis of our discussion (see the last Section)
of the `plateau' in $P(T)$ for values of $\alpha_o$ near $-1.5$. The
presence of the `plateau' is independent of the halo model; it is a
robust property carried over from $P_o(T)$ to $P(T)$. (Our results
for various values of the `true' halo parameters, $\gamma$, $\sigma_o$, 
$\sigma_+$, $r_o$ and $l$, indicate  
that $\alpha$ and $\beta$ are generally insensitive to the halo model
uncertainties in the above range of $\alpha$.)

 By contrast,
for $\alpha_o \agt 0.5$ the shift is significant.  
In this range of $\alpha_o$,  $\Psi ({\bf c},{\bf c}_o)$ as a function of
$\alpha$ and $\beta$ is very flat and insensitive to shifts in the
two parameters.  Due to this circumstance, even a slight `tilt' to
the function $\Psi ({\bf c},{\bf c}_o)$ due to a switch to another
halo model is likely to lead to a large shift in the position
of the function's maximum in the $(\alpha, \beta)$ plane.  This is closely
related to the basic reasons for large statistical errors for $\alpha_o \agt 0.5$
that we discussed in the last section. 
 From the point of view of the
inference of $\alpha$ and $\beta$, the bias due to  the uncertainty of
the halo model would in this case only compound  
already very large statistical
errors.  The same holds for  large negative values of $\alpha$ ($ < -2.5$).

In addition to the effect on the mass function determination,
MHO distribution and kinematics uncertainties will bear on the
determination of the density and total mass in the massive objects'
halo.  If $\varepsilon =$ const. the two integrations in the
total rate\ (\ref{rate.y}) can be performed in the reversed order
\begin{eqnarray}
\label{recerse.int}
&&\hspace{-0.3cm}\int^{\infty}_{0}dT\; T^{2(\alpha + 1)} 
               \int_{\mu_o \frac{10^{-\beta/2}}{T^2}}^
          {\mu_o \frac{10^{\beta/2}}{T^2}}dy\; y^{\alpha} F(y) \nonumber \\
	  \nonumber \\	  
&&\hspace{0.2cm}= \int^{\infty}_{0} dy\; y^{\alpha}F(y) 
               \int^{\sqrt{\mu_o/y}\, 10^{\beta/4}}
        _{\sqrt{\mu_o/y}\, 10^{-\beta/4}} dT\; T^{2(\alpha +1)} \nonumber \\
	\nonumber \\
&&\hspace{0.2cm}=\frac{1}{2}\mu_o ^{\alpha +3/2} C_{\beta}(\alpha + 1/2)
  \int_0 ^{\infty}F(y) y^{-3/2} dy,
\end{eqnarray}
and the total rate takes on the form 
\begin{equation}
\label{rate.fin}
\Gamma = K \rho_o \frac{1}{\bar{\mu}} \Xi (\alpha, \beta)
 \int_0 ^{\infty}F(y) y^{-3/2} dy,
\end{equation}
where $K$ is a constant independent of either the halo model or
the MHO mass function and
\begin{equation}
\label{ksi.def}
\Xi (\alpha,\beta) \equiv\frac{C_{\beta}(\alpha + 1/2)}{\sqrt{C_\beta (\alpha)
 C_{\beta}(\alpha + 1)}}.
\end{equation} 
Obviously [see equation\ (\ref{PmuT})], $F(y)$  depends 
only on the halo model but not on the parameters
of the mass function.  It also turns out that the equality 
$\Xi (\alpha,\beta)/\Xi (\alpha_o ,\beta_o ) = 1$
[$\alpha (\alpha_o , \beta_o)$ and $\beta  (\alpha_o , \beta_o)$ are again 
the best fits obtained with the `false' halo model] 
is satisfied with better than 1\% accuracy for the range of $\alpha$ and $\beta$
discussed in this paper.  Since the inferred $\bar{\mu}$, shown
in Fig.\ 9, depends only very weakly on $\alpha$ and $\beta$ it follows
that the ratio $(\rho_o)_{\rm biased}/(\rho_o)_{\rm real}$ is practically
independent of $\alpha$ and $\beta$.  

In the case discussed in this section so far, where the CS halo is
chosen as the `real' one and the SIS as the halo model used in the inference
of the mass function, we obtain $(\rho_o)_{\rm biased}/(\rho_o)_{\rm real} \simeq 0.5$.
Although the inferred value of the local MHO halo density $\rho_o$
is only half the actual one, the much slower fall-off of the SIS density at
larger distances leads to the inferred total halo mass, $M_{\rm tot}=
4\pi \int_{R_{\odot}}^{R_{\rm LMC}} \rho_o (r/R_{\odot})^{-\gamma}r^2 dr$, between 
the Sun and the LMC being about twice the actual amount.

Since both the isothermal sphere
and the `concentrated' sphere can be viewed as plausible models of
the massive objects' halo,  the  parameter shifts evaluated so far
in this section provide an estimate of the magnitude of error due
to halo structure uncertainty.
As we have stated above,
due to
the difficulty of distinguishing among different halo structures
on the basis of event duration measurements only,  
it would appear unlikely that these (systematic) errors should be
possible to eliminate.

\begin{figure}
\centering
\centerline{\epsfxsize= 8 cm \epsfbox[70 50 560 550]{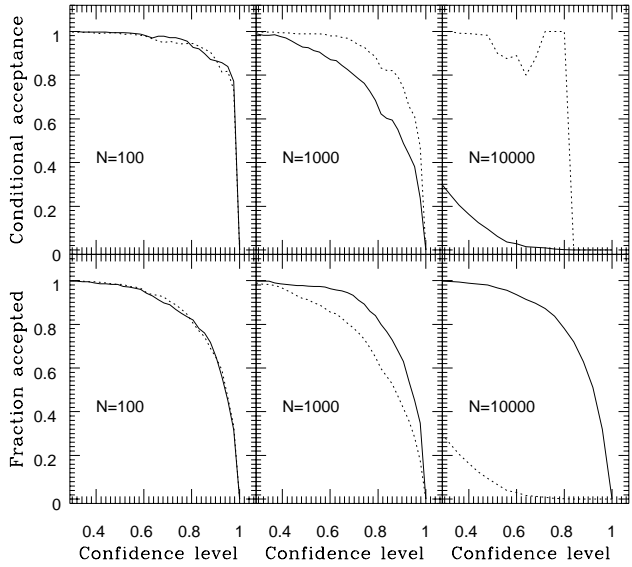}}
\caption{The acceptable fraction (bottom) of best fit curves $P(T)$ based on the
`real' CS halo (solid line) and `false' SIS halo (dottes line) as
functions of Kolmogorov-Smirnov confidence levels for $N=100$,
$N=1000$
and $N=10000$ detected events. The `real' mass function is $\bar{\mu} =0.4$,
$\alpha = -2$ and $\beta = 2$.  In the upper
panels the solid lines give the `probability' that {\it if}
a CS-based best fit is accepted at the given confidence level, the
SIS-based best fit will also pass the Kolmogorov-Smirnov test.
The dotted lines show the probability of acceptance of
a CS-based fit given that the SIS-based fit is acceptable.  
} 
\end{figure}

\begin{figure}
\centering
\centerline{\epsfxsize= 8 cm \epsfbox[60 40 560 755]{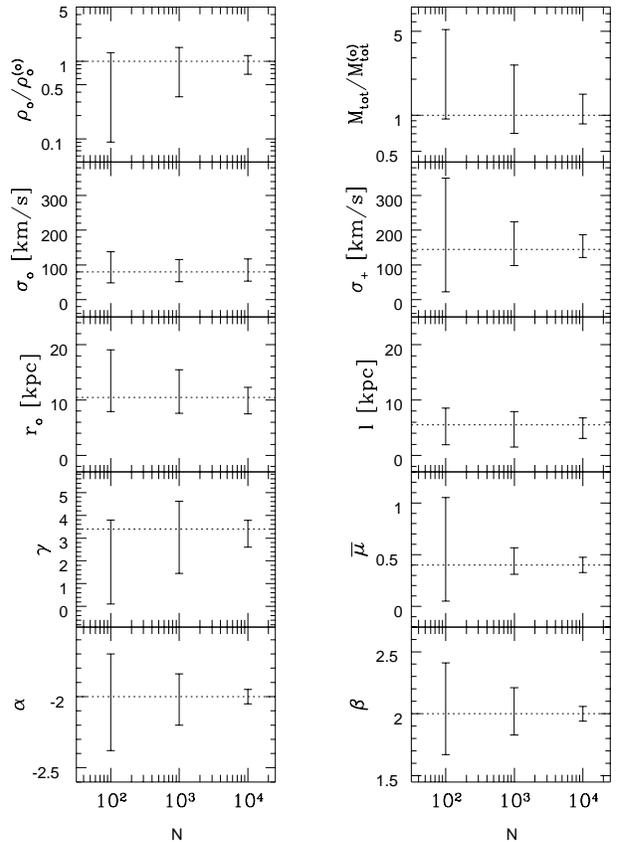}}
\caption{
 `$1\sigma $' confidence intervals (obtained by simulations where {\it all}
eight parameters are varied in the maximum likelihood method) as functions
of the number of detected events $N$.  The
dotted lines give the values of the `real' parameters.
}
\end{figure}

On the other hand, we have noticed in the early part of this section that
for $\alpha$ in the range most conducive to mass function parameter inference ($\alpha
\sim -1.5$), there is a residual difference between the differential event rate
based on one halo model and its best match based on a different halo structure
(upper panel of Fig. 8).  Can this difference be exploited in a realistic
statistical inference in order to reduce the effect of the halo structure
uncertainty?
Specifically, with how many detected events do we need to sample the 
differential rate so that the small difference can be recognised?

In order to answer this question we perform a  number (400)
of Monte-Carlo simulations of 
the mass function parameters' statistical inference.
 Again, we assume that the `real' halo is CS, $\bar{\mu}_o =0.4$, $\alpha =-2$, 
 and $\beta=2$,
generate a given number of events ($N=100$, $1000$ or $10000$) and obtain maximum
likelihood fits based in turns on the CS and the (`false') SIS halo.  In Fig. 10
we show the results of the  Kolmogorov-Smirnov test of  acceptability of
the maximum likelihood fits (based on the two different halos) as 
models for
the observed data.  

At $N=100$ it makes virtually no difference which
halo model is used as a basis for statistical inference:  the
small discrepancy between the two event rate curves observed in Fig. 8 is 
completely swamped by the statistical noise.  On the other
hand, for $N=1000$ the difference starts to be significant, although
the probability of confusion between the two halos in
any single data realisation is still high.  At $N=10000$ the ambiguity
is completely resolved. Obviously, it
is at about $N=1000$ that we may first start discerning
the halo structure effects and thus hope to be able to reduce the 
(halo model-induced) errors in the
mass function parameters estimated earlier in this section.

This conclusion is borne out by Monte-Carlo simulations where 
all eight parameters ($\gamma$, $\sigma_o$, $\sigma_+$, $r_o$, $l$,
$\bar{\mu}$, $\alpha$ and $\beta$) are varied with a uniform prior
to obtain maximum-likelihood
fits to events distributed according to the differential event rate whose
parameters are indicated by the dotted lines in Fig. 11.  The vertical bars
show `$1\sigma$' intervals for inferred values of the parameters. [The inferred
$\rho_o$ and $M_{\rm tot}$ are simply read out from the event rate (\ref{rate})
given a fixed event number of events and observation time without
allowing for the inevitable Poisson count noise.]

As we concluded earlier in this section, for $\alpha_o = -2$
the effects of the halo uncertainty on $\alpha$ and $\beta$ are
rather weak and these parameters are obtained with  a
 good accuracy from between 100
and 1000 events. The halo model uncertainty adds little to the errors
estimated in the last section.  On the other hand, the error in
 $\bar{\mu}$ falls bellow the level
estimated above in this section only at $N\sim 1000$.  Notice that
at $N=10000$ the error is larger by a factor of about 10 than the value obtained
with a fixed halo model (see Fig.\ 5).

The improvement at $N\approx 1000$ is not necessarily reflected in
an accurate knowledge of the structure of the halo: for instance, 
$r_o$ and $l$, and thus the velocity profile, are not accurately
determined even for the largest $N$'s shown in Fig. 11.  It is rather that
for $N$ near $1000$ the halo is constrained just enough to allow
a more reliable mass function inference; the
small gap between the two curves of the upper panel of Fig. 8 can thus be
much reduced with a rather broad range of best-fit halo models.

Probably the most important quantities characterising the halo,
its local density $\rho_o$ and total mass between the Sun and LMC, $M_{\rm tot}$
(themselves functions of the above eight parameters and the observed event rate)
clearly gain in accuracy with increasing $N$ (see the two top panels of Fig. 11).
However, it should be pointed out that only at around $N=10000$ is the
error due to halo structure uncertainty reduced to roughly the level of
the simple Poisson count fluctuation $1/\sqrt{6} \approx 40\%$ associated with
the total number of (reliable)
LMC events detected by the time of this writing.

\subsection*{Acknowledgements}
This research was generously supported by Danmarks Grundforskningsfond
through its establishment of the Theoretical Astrophysics Center.

\appendix
\section*{Appendix: Bias, error and the maximum likelihood method}

In this appendix we summarise some known mathematical results pertinent to the
problem of estimating parameters of a probability distribution, given a
set of measurements [see also \cite{cramer}].  

Suppose we have a set of $N$ measurement results
${\bf x} \equiv \{x_{1}, x_{2}$, ..., $x_{N}\}$. 
Suppose also that the measurements are
distributed  according to the
probability distribution $f(x|{\bf c})$  ($\int f dx = 1$),
where ${\bf c}$ denotes a set
of $p$ parameters which are {\it  a  priori} unknown.
{\it Estimator} $\hat{c}({\bf x})$ is then
a function of the measurement results
that can serve as a reasonable estimate of
 the values of the underlying
parameters ${\bf c}$.

Any statistical inference is, in principle, subject to error and bias in the
inferred parameters. For each parameter $c_{\mu}$ the variance of estimation
is defined as
\begin{equation}
\label{sigma.def}
\sigma^{2}_{c_{\mu}} \equiv E(\hat{c}^{2}_{\mu}) -
                        E^2 (\hat{c}_{\mu}),
\end{equation}
where $E$ denotes the expectation value of an ensemble whose each member
consists of $N$ measurements under the same conditions; 
$\hat{c}_{\mu}$ are
the values of parameters estimated from the n measurements in a member of
the ensemble. On the other hand, the bias
\begin{equation}
\label{bias.def}
b_{\mu} \equiv E(\delta c_{\mu}) = E(\hat{c}_{\mu}) - c_{\mu}
\end{equation}
is the systematic departure of estimated parameters from the
`true' parameters ${\bf c}$.

Given the dependence of bias on the the underlying parameters,
the errors of estimation are bounded from below by the
Cramer (also known as Frechet-Cramer-Rao) limit, as we
derive in the following.

Denoting $F({\bf x}|{\bf c})\equiv f(x_1 |{\bf c})f(x_2 |{\bf c})\cdots 
f(x_{N}|{\bf c})$,  the definition of bias (\ref{bias.def})
can be rewritten as
\begin{equation}
\label{bias.def1}
\int \hat{c}_{\mu}({\bf x}) F({\bf x}|{\bf c})d^n {\bf x} = c_\mu + b_\mu.
\end{equation}
Differentiating the above equation with respect to parameter
$c_{\nu}$ we obtain
\begin{equation}
\label{diff.bias}
E(\hat{c}_\mu l_{,\nu}) = E\left[\left(\hat{c}_\mu - E(\hat{c}_\mu)\right)
   l_{,\nu}\right]
 = \delta_{\mu \nu} + b_{\mu,\nu},
\end{equation}
where we have used
\begin{equation}
\label{l.def} 
l({\bf x},{\bf c}) \equiv \ln F({\bf x}|{\bf c}),
\end{equation}
and the fact
that
\begin{equation}
\label{exp.l}
E(l_{,\mu}) = N\int [\ln f(x |{\bf c})]_{,\mu} f(x |{\bf c}) dx =0,
\end{equation}
due to the normalisation of $f$.
Convolving the matrix equation (\ref{diff.bias}) with
arbitrary $p$-dimensional vectors $u_\mu$ and $v_\nu$ and
using the Schwartz inequality, we obtain
\begin{equation}
\label{schwartz}
V_{\mu\nu}u_\mu u_\nu \; I^{(n)}_{\rho\sigma}v_\rho v_\sigma
\geq \left( u_\mu v_\mu + b_{\mu ,\nu}u_{\mu} v_\nu \right)^2,
\end{equation}
where we have used Einstein's convention of
summation over repeated indices and defined the variance matrix
\begin{equation}
\label{V.def}
V_{\mu\nu}\equiv E\left[\left(\hat{c}_\mu -E(\hat{c}_\mu)\right)
       \left(\hat{c}_\nu -E(\hat{c}_\nu)\right)\right],
\end{equation}
and the information matrix
\begin{eqnarray}
\label{I.def}
I^{(n)}_{\mu\nu} &=&E(l_{,\mu}l_{,\nu}) = N I_{\mu\nu} 
               \nonumber \\ &=&
  N\int \left(\ln f(x|{\bf c})\right)_{,\mu}
        \left(\ln f(x|{\bf c})\right)_{,\nu} 
    f(x|{\bf c}) \; dx.
\end{eqnarray}

It is convenient at this point  to simplify notation 
by introducing Dirac-style bra/ket notation. Thus, $|u>$
denotes the column $u_\mu$ while $<u|$ denotes the corresponding
row. The equation (\ref{schwartz}) can now be rewritten as
\begin{equation}
\label{schwartz.1}
<u\,|V|u> \hspace{0.2cm} \geq \hspace{0.2cm} 
         \frac{\left(<u\,|v> + <u\,|B|v>\right)^2}{<v\,|I|v>},
\end{equation}
where $B_{\mu\nu} \equiv b_{\mu ,\nu}$.
Since $|v>$ is arbitrary, we can find the maximum of the
right-hand side in the above equation (for a given $|u>$) if we 
require that the first derivative with respect to $|v>$ be
zero. This condition leads to
\begin{equation}
\label{max.condition}
<u|({\bf 1} + B)\left({\bf 1} - \frac{|v><v|}{<v| I |v>} I\right) = 0.
\end{equation}
It is straightforward to see that the vector satisfying condition
(\ref{max.condition}) is
\begin{equation}
\label{max.solution}
|v> \; = I^{-1} \left({\bf 1} + B^{T}\right)|u>,
\end{equation}
which by substitution in equation (\ref{schwartz.1})
gives the Cramer inequality
\begin{equation}
<u|V|u>\hspace{0.2cm} \geq \hspace{0.2cm}
 <u| \left({\bf 1} + B\right) I^{-1} \left({\bf 1} + B^{T}\right)|u>,
\end{equation}
where $|u>$ is an arbitrary $p$-dimensional vector.

In general, the Cramer inequality is only of limited utility:
the actual estimator we are using may have variance that 
far exceeds the Cramer limit.  In addition, it may be difficult
to determine the bias, which is needed to compute the Cramer limit
in the first place.
However, as we show below, the method of maximum likelihood
is asymptotically (for $N$ large enough) unbiased and
approaches the Cramer limit.

In the maximum likelihood method one adopts as
estimates $\hat{\bf c}$ of the parameters determining
distribution $f(x|{\bf c})$ those values of the parameters
that yield the maximum of $l({\bf x}| {\bf c})$ 
[ see equation (\ref{l.def})] given a set ${\bf x}$
of $N$ measurements. Expanding the maximum likelihood
condition in the vicinity of the true values of the parameters
${\bf c}$, $0 = l_{,\mu} (\hat{\bf c}) = l_{,\mu}({\bf c}) +
(\hat{c}_\nu - c_\nu )l_{,\mu\nu}({\bf c})$ we obtain
the `shift' from the true parameters
\begin{equation}
\label{ml.shift}
\hat{c}_\mu - c_\mu = -\left[ l_{,\mu\nu}\right]^{-1} l_{,\nu} ,
\end{equation}
where $\left[ l_{,\mu\nu}\right]^{-1}$ denotes the inverse
matrix of $\left[ l_{,\mu\nu}\right]$.
For $n$ large enough, $\left[ l_{,\mu\nu}\right]^{-1}$ can
be approximated by its ensemble average 
\begin{equation}
\label{asymptl.1}
\left[ l_{,\mu\nu}\right]^{-1} \approx 
E\left\{\left[ l_{,\mu\nu}\right]^{-1}\right\}\approx
\left[ E\left(l_{,\mu\nu}\right) \right]^{-1} =
\left[ I^{(N)}_{\mu\nu}({\bf c}) \right]^{-1},
\end{equation}
where we have substituted $E(l_{,\mu\nu}) = - E(l_{,\mu} l_{,\nu})$,
which is a straightforward consequence of the normalisation of $f$.
Due to equation (\ref{exp.l}) the bias asymptotically vanishes,
 while the variance (\ref{V.def}) approaches
the Cramer limit
\begin{eqnarray}
\label{asympt.var}
V_{\mu\nu} &=& E\left\{ \left[ l_{,\mu\rho}\right]^{-1}
	\left[ l_{,\nu\sigma}\right]^{-1}l_\rho l_\sigma \right\}
	 \nonumber \\
	&\approx & \left[ I^{(N)}_{\mu\rho}({\bf c}) \right]^{-1}
	\left[ I^{(N)}_{\nu\sigma}({\bf c}) \right]^{-1}
	E\left[l_\rho l_\sigma\right]
	\nonumber \\
	&\approx & \left[ I^{(N)}_{\mu\nu}({\bf c}) \right]^{-1},
\end{eqnarray}
where, in the last equality we used equation (\ref{I.def}).
We denote the corresponding `Cramer'
deviation limit as $\Delta_{\rm c} c_{\mu} \equiv 
(n V_{\mu\mu})^{1/2}$.  

In the above otline of the maximum likelihood method we
assumed that all uncertainty regarding the underlying
distribution function $f({\bf x}| {\bf c})$ stems
from the uncertainty of the values of a well
defined, finite set of parameters, while the functional
dependence $f({\bf x}| {\bf c})$ is known.
In the more general case, the functional dependence 
is not known accurately. We can then ask the following
question: if we happen to be using a `wrong' functional
dependence $f({\bf x}| {\bf c}) = 
f_o ({\bf x}| {\bf c}) + \delta f({\bf x}| {\bf c})$ instead
of the correct function $f_o ({\bf x}| {\bf c})$,
what will be the bias (\ref{bias.def}), i.e., the
average departure from the true parameters.  Obviously,
in this case the bias need not be asymptotically zero.

More specifically, the maximum likelihood method looks for the
maximum of  $l({x}|\hat{\bf c}) \equiv 
\Sigma_{i=0}^{N} \ln f(x_i |\hat{\bf c})$, 
where ${\bf x}$ are results of measurements distributed 
according to the true function $f_o$, i.e., it seeks
the values $\hat{\bf c}$ for which $l({x}|\hat{\bf c})_{,\mu} =0$.
A reasonable approach to estimating the average value of $\hat{\bf c}$
that this method would give for large $n$, 
may be to look for those values
of $\hat{\bf c}$ that maximize
\begin{equation}
\label{Psi.def}
\Psi (\hat{\bf c},{\bf c}) = \int f_o (x | {\bf c}) 
          \ln f( x |\hat{\bf c}) d x.
\end{equation} 
In other words, the requirement is 
\begin{equation}
\label{Psi.max}
 \partial\Psi (\hat{\bf c},{\bf c})
/\partial\hat{c}_\mu = 0.
\end{equation}
Note that if $f =f_o$, this condition simply implies $\hat{\bf c}
= {\bf c}$.

Unfortunately, this prescription  finds the maximum
of the average of $l({\bf x}|\hat{\bf c})$ which 
quite obviously need not be the same as
the required average of those $\hat{\bf c}$ that maximize
$l({\bf x}|\hat{\bf c})$ for different measurement sets 
${\bf x}$.
Still, for small enough $\delta f$ the two quantities are
very close to each other.  Indeed, in this case 
\begin{eqnarray}
\label{l.wrong}
l({x}|\hat{\bf c}) 
 &=& \Sigma_{i=0}^{N}\ln [f_o (x_i |\hat{\bf c}) + 
  \delta f(x_i |\hat{\bf c})] \nonumber \\
   &\approx & \Sigma_{i=0}^{N} \left(\ln f_{o} + \delta f/f_o\right).
\end{eqnarray}
From equation (\ref{ml.shift}) we then obtain
\begin{equation}
\label{c.bias1}
b_\mu = E(\hat{c}_\mu - c_\mu) \approx \left[I_{\mu\nu}\right]^{-1}
    E\left(
   \frac{1}{f_o} \delta f_{,\nu}  -  \frac{\delta f}{f_o^2} f_{o,\nu}
    \right),
\end{equation}
where we have kept only terms linear in $\delta f$ and
used the fact that $E(l_o) =0$.  On the other hand,
the condition 
$\partial\Psi (\hat{\bf c},{\bf c})/\partial\hat{c}_\mu = 0$
gives
\begin{eqnarray}
\label{c.bias2}
0 \hspace{-0.1cm}
  &=& \int f_o dx\, (x |{\bf c}) \frac{\partial}{\partial \hat{c}_\mu}
       \left[ \ln f_o (x |\hat{\bf c}) + \frac{\delta f}{f_o}\right]
       \nonumber \\
  &\approx &
  \int f_o dx\, \bigg[(\ln f_o (x |{\bf c}) )_{,\mu}
   + (\hat{c}_\nu - c_{\nu} )(\ln f_o (x |{\bf c}) )_{,\mu\nu}
   \nonumber \\
    & & \hspace{3cm}
   +\; \frac{1}{f_o} \delta f_{,\nu}  -  \frac{\delta f}{f_o^2} f_{o,\nu}
    \bigg],
\end{eqnarray}
which in the limit of large $n$ leads to equation
(\ref{c.bias1}).

Although, strictly speaking, condition (\ref{Psi.max}) holds
only for small $\delta f$, its cautious application can yield
a reasonable estimate of bias even away from this strict limit.

\end{document}